\documentclass[12pt]{article}
\usepackage{lscape}
\usepackage{citesort}
\usepackage{amssymb}
\usepackage{appendix}
\usepackage{multirow}

\begin{document}
\def\qq{\langle \bar q q \rangle}
\def\uu{\langle \bar u u \rangle}
\def\dd{\langle \bar d d \rangle}
\def\sp{\langle \bar s s \rangle}
\def\GG{\langle g_s^2 G^2 \rangle}
\def\Tr{\mbox{Tr}}
\def\figt#1#2#3{
        \begin{figure}
        $\left. \right.$
        \vspace*{-2cm}
        \begin{center}
        \includegraphics[width=10cm]{#1}
        \end{center}
        \vspace*{-0.2cm}
        \caption{#3}
        \label{#2}
        \end{figure}
	}
	
\def\figb#1#2#3{
        \begin{figure}
        $\left. \right.$
        \vspace*{-1cm}
        \begin{center}
        \includegraphics[width=10cm]{#1}
        \end{center}
        \vspace*{-0.2cm}
        \caption{#3}
        \label{#2}
        \end{figure}
                }

\def\ds{\displaystyle}
\def\beq{\begin{equation}}
\def\eeq{\end{equation}}
\def\bea{\begin{eqnarray}}
\def\eea{\end{eqnarray}}
\def\beeq{\begin{eqnarray}}
\def\eeeq{\end{eqnarray}}
\def\ve{\vert}
\def\vel{\left|}
\def\ver{\right|}
\def\nnb{\nonumber}
\def\ga{\left(}
\def\dr{\right)}
\def\aga{\left\{}
\def\adr{\right\}}
\def\lla{\left<}
\def\rra{\right>}
\def\rar{\rightarrow}
\def\lrar{\leftrightarrow}  
\def\nnb{\nonumber}
\def\la{\langle}
\def\ra{\rangle}
\def\ba{\begin{array}}
\def\ea{\end{array}}
\def\tr{\mbox{Tr}}
\def\ssp{{\Sigma^{*+}}}
\def\sso{{\Sigma^{*0}}}
\def\ssm{{\Sigma^{*-}}}
\def\xis0{{\Xi^{*0}}}
\def\xism{{\Xi^{*-}}}
\def\qs{\la \bar s s \ra}
\def\qu{\la \bar u u \ra}
\def\qd{\la \bar d d \ra}
\def\qq{\la \bar q q \ra}
\def\gGgG{\la g^2 G^2 \ra}
\def\GG{\langle g_s^2 G^2 \rangle}
\def\g5{\gamma_5 \not\!q}
\def\x{\gamma_5 \not\!x}
\def\g5{\gamma_5}
\def\sb{S_Q^{cf}}
\def\sd{S_d^{be}}
\def\su{S_u^{ad}}
\def\sbp{{S}_Q^{'cf}}
\def\sdp{{S}_d^{'be}}
\def\sup{{S}_u^{'ad}}
\def\ssp{{S}_s^{'??}}

\def\sig{\sigma_{\mu \nu} \gamma_5 p^\mu q^\nu}
\def\fo{f_0(\frac{s_0}{M^2})}
\def\ffi{f_1(\frac{s_0}{M^2})}
\def\fii{f_2(\frac{s_0}{M^2})}
\def\O{{\cal O}}
\def\sl{{\Sigma^0 \Lambda}}
\def\es{\!\!\! &=& \!\!\!}
\def\ap{\!\!\! &\approx& \!\!\!}
\def\ar{&+& \!\!\!}
\def\arrr{\!\!\!\! &+& \!\!\!}
\def\ek{&-& \!\!\!}
\def\vev{&\vert& \!\!\!}
\def\kek{\!\!\!\!&-& \!\!\!}
\def\cp{&\times& \!\!\!}
\def\se{\!\!\! &\simeq& \!\!\!}
\def\eqv{&\equiv& \!\!\!}
\def\kpm{&\pm& \!\!\!}
\def\kmp{&\mp& \!\!\!}
\def\mcdot{\!\cdot\!}
\def\erar{&\rightarrow&}


\def\simlt{\stackrel{<}{{}_\sim}}
\def\simgt{\stackrel{>}{{}_\sim}}


\title{
         {\Large
                 {\bf
Magnetic moments of heavy $J^P = {1\over 2}^+$ baryons in 
light cone QCD sum rules
                 }
         }
      }

\author{\vspace{1cm}\\
{\small T. M. Aliev \thanks {e-mail:
taliev@metu.edu.tr}~\footnote{permanent address:Institute of
Physics,Baku,Azerbaijan}\,\,, T. Barakat \thanks {e-mail:
tbarakat@KSU.EDU.SA}\,\,, M. Savc{\i} \thanks
{e-mail: savci@metu.edu.tr}} \\
{\small Physics Department, Middle East Technical University,
06531 Ankara, Turkey }\\
{\small $^\ddag$ Physics and Astronomy Department, King Saud University, Saudi Arabia}}

\date{}

\begin{titlepage}
\maketitle
\thispagestyle{empty}

\begin{abstract}

The magnetic moments of heavy sextet $J^P = {1\over 2}^+$ baryons are
calculated in framework of the light cone QCD sum rules method. Linearly
independent relations among the magnetic moments of these baryons are
obtained. The results for the magnetic moments of heavy baryons obtained in
this work are compared with the predictions of the other approaches.

\end{abstract}

~~~PACS numbers: 11.55.Hx, 13.40.Em, 14.20.Mr, 14.20.Lq

\end{titlepage}

\section{Introduction}

The quark model predicts the existence of heavy baryons composed of single,
double and triple quarks. Essential improvement has been achieved in heavy
baryon spectroscopy in resent years. All baryons with a single charm quark
that are predicted by the quark model have been observed in experiments.
Moreover, heavy baryons with a single bottom quark, such as $\Lambda_b$,
$\Sigma_b$, $\Xi_b$ and $\Omega_b$ have also been discovered (for a review
see \cite{Rnil01}).

These progresses in experiments stimulated future investigation for the
properties of these baryons at LHC, as well as further theoretical studies    
on this subject.

Remarkable information about the internal structure of baryons can be gained  
by studying their magnetic moments. Magnetic moments of the heavy baryons
have long been under the focus of theoretical physicists, and they have been
studied in framework of various approaches. So far, magnetic moments of charmed
heavy baryons have been calculated in framework of the naive quark model
\cite{Rnil02,Rnil03}, relativistic quark model \cite{Rnil04}, chiral
perturbation model \cite{Rnil05}, hypercentral model \cite{Rnil06}, soliton
model \cite{Rnil07}, skyrmion model \cite{Rnil08}, bag model
\cite{Rnil09}, QCD sum rules method \cite{Rnil10},
nonrelativistic quark model \cite{Rnil11}, phenomenological relativistic
model \cite{Rnil12}, quark model with confinement law potential
\cite{Rnil13}, and chiral constituent quark model, respectively.

Magnetic moments of heavy hadrons have also been studied in numerous
works within the QCD sum rules method. Magnetic moments of the $\Lambda_c$
and $\Sigma_c$ baryons have been calculated in framework of the traditional
QCD sum rules method in \cite{Rnil10}; of the $\Lambda_Q$ and $\Xi_Q$ $(Q=c$
or $b)$ baryons within the light version of the QCD sum rules method in
\cite{Rnil15,Rnil16}, respectively. It should be noted here that magnetic
moments of the spin-3/2 heavy baryons have already been analyzed within the
same approach in \cite{Rnil17}.

The goal of the present work is calculation of the magnetic moments of heavy
$\Sigma_Q$, $\Xi_Q^\prime$ and $\Omega_Q$ baryons within the light cone QCD
sum rules method.

The paper is organized as follows. In section 2, sum rules for the
magnetic moments of the above-mentioned baryons are constructed. Numerical
results of our calculations are presented in section 3. This section further
contains comparison of our results with the predictions of other approaches,
and concluding remarks.

\section{Light cone sum rules for the sextet heavy baryons}

We start this section by giving a brief summarizing of the classification of
heavy baryons in SU(3) symmetry. According to SU(3) symmetry, heavy baryons
with single heavy quark belong to either symmetric sextet or antisymmetric
anti-triplet flavor representations. As has already been noted , magnetic
moments of the $\Lambda_Q$ and $\Xi_Q$ baryons belonging to anti-symmetric
flavor representations of SU(3) are calculated within the light QCD sum rules 
method in \cite{Rnil15} and \cite{Rnil16}, respectively. In the present work
we calculate magnetic moments of the heavy baryons belonging to sextet
representation of SU(3) group.

In order to calculate magnetic moments of the heavy sextet
baryons, we start by considering the following correlation function:
\bea
\label{enil01}
\Pi (p,q) = i \int d^4x e^{ipx} \lla \gamma(q) \vel \mbox{\rm T} \{
\eta_Q (x) \bar{\eta}_Q(0) \}\ver 0 \rra ~,
\eea
where $\eta_Q$ is the interpolating current of the heavy spin-1/2 baryon.
The main task in constructing the sum rules for magnetic moments of the heavy
sextet baryons is the calculation of the correlation function in terms of
the photon distribution amplitude by using the operator product expansion
(OPE) over twist from one side, and in terms of the hadrons from the other
side, and then equating both representations. Calculation of the correlation
function from the hadronic side is accomplished by inserting a complete set of
hadrons carrying the same quantum numbers of the interpolating current
$\eta_Q$. Isolating the the ground state's contribution, we obtain,
\bea
\label{enil02}
\Pi = {\la 0 \ve \eta_Q \ve B_Q (p_2)
\ra \over p_2^2-m_{B_Q}^2} \la B(p_2) \gamma (q) \ve
B_Q (p_1)  \ra_\gamma { \la B_Q (p_1) \ve \bar{\eta}_Q \ve 0 \ra
\over p_1^2-m_{B_Q}^2} + \cdots,
\eea
where dots correspond to the contributions of higher states and continuum,
$p_!=p_2+q$. In further analysis we shall make the replacement $p_2=p$. The
matrix elements in Eq. (\ref{enil02}) are determined as,
\bea
\label{enil03}
\la 0 \vel \eta_Q \ver B_Q (p) \ra \es \lambda_Q u_{B_Q}(p)~, \\
\label{enil04}
\la B_Q(p) \ve B_Q (p_1) \ra_\gamma \es \varepsilon^\mu
\bar{u}_{B_Q} (p) \Bigg[ f_1 \gamma_\mu -f_2 {i \sigma_{\mu\nu} \over 2 m_{B_Q}}
q^\nu \Bigg] u_{B_Q}(p_1)~,
\eea
where $\varepsilon^\mu$ is the photon polarization vector, $f_1$ and $f_2$
are the form factors. Using the equation of motion, Eq. (\ref{enil04}) can
be written as,
\bea
\label{enil05}
\la B_Q(p) \ve B_Q (p_1) \ra_\gamma = \bar{u}_{B_Q} (p) \Bigg[(f_1+f_2)
\gamma_\mu + {(p+p_1)_\mu \over 2 m_{B_Q}} \Bigg] u_{B_Q} (p_1)
\varepsilon^\mu~.
\eea
Since the photon involved in these transitions is a real photon , we set $q^2=0$,
and hence in calculation of the magnetic moments of the heavy sextet baryons  
the values of the form factors are needed only at $q^2=0$.  

Substituting Eqs. (\ref{enil03}) and (\ref{enil05}) in Eq. (\ref{enil02}),
and performing summations over spins of the heavy baryons, for the hadronic
of the correlation function we get,
\bea
\label{enil06}
\Pi = \lambda_{B_Q}^2 \varepsilon^\mu {(\rlap/{p} + m_{B_Q})\over p^2 -
m_{B_Q}^2} \Bigg[(f_1+f_2) \gamma_\mu + {(p+p_1)_\mu \over 2 m_{B_Q}} \Bigg]
{(\rlap/{p}_1 + m_{B_Q})\over p_1^2 - m_{B_Q}^2}~.
\eea
We observe from Eq. (\ref{enil06}) that the correlation function contains
many structures, any of them can be chosen in calculating magnetic
moments of the sextet baryons, and in this respect we choose the structure  
$\rlap/{p}\rlap/{\varepsilon}\rlap/{q}$. This structure envelopes the magnetic form
factor $f_1+f_2$, and at $q^2=0$ it
gives the magnetic moment $\mu_{B_Q}$ of the heavy baryons in units of
$e\hbar/2 m_{B_Q}$. As a result, the correlation function can be written in
terms of the magnetic moment of heavy baryons as,
\bea
\label{enil07}
\Pi = \lambda_{B_Q}^2 {1\over m_{B_Q}^2 - p^2} \mu_{B_Q} {1\over m_{B_Q}^2 -
(p+q)^2}~.
\eea

In order to calculate the correlation function in terms of quark and gluon
degrees of freedom and photon distribution amplitudes, the expressions of
the interpolating currents of the heavy baryons are needed. The general form
of the interpolating currents of the heavy spin-1/2 positive parity baryons 
is given as (see for example \cite{Rnil18}),
\bea
\label{enil08}
\eta_{B_Q} \es - {1\over \sqrt{2}} \epsilon^{abc} \Big\{ (q_1^{aT} C Q^b)
\gamma_5 q_2^c + t (q_1^{aT} C \gamma_5 Q^b) q_2^c -
(Q^{aT} C q_2^b ) \gamma_5 q_1^c - t (Q^{aT} C \gamma_5 q_2^b ) q_1^c
\Big\}~,
\eea
where $a,b,c$ are the color indices, $C$ is the charge conjugation operator,
and $t$ is an arbitrary parameter whose value at $t=-1$ gives the so-called
Ioffe current. The quark contents of the sextet heavy sextet baryons are
given in Table 1.   


\begin{table}[h]

\renewcommand{\arraystretch}{1.3}
\addtolength{\arraycolsep}{-0.5pt}
\small
$$
\begin{array}{|c|c|c|c|c|c|c|}
\hline \hline   
 & \Sigma_{b(c)}^{+(++)} & \Sigma_{b(c)}^{0(+)} & \Sigma_{b(c)}^{-(0)} &
             \Xi_{b(c)}^{\prime -(0)}    & \Xi_{b(c)}^{\prime 0(+)}  & 
\Omega_{b(c)}^{-(0)} \\
\hline \hline
q_1   & u & u & d & d & u & s \\
q_2   & u & d & d & s & s & s \\
\hline \hline
\end{array}
$$
\caption{Quark contents of the heavy sextet baryons.}
\renewcommand{\arraystretch}{1}
\addtolength{\arraycolsep}{-1.0pt}
\end{table}   

Using the expression for the interpolating current and Wick's theorem,
the theoretical part of the correlation function for the
$\Sigma_b^0$ can be written as,
\bea
\label{enil09}
\Pi^{\Sigma_b^0} \es
         -3 \Big\{ \gamma_5 S_d(x) S_b^\prime(x) S_u(x)  
         \gamma_5 + \gamma_5 S_u(x) S_b^\prime(x) 
         S_d(x) \gamma_5 + 
       \gamma_5 S_u(x) \gamma_5 
        Tr \left[S_b(x) S_d^\prime(x) \right] \nnb \\
\ar
       \gamma_5 S_d(x) \gamma_5 
        Tr \left[S_u(x) S_b^\prime(x) \right] +
      t \Big( \gamma_5 S_d(x) \gamma_5 S_b^\prime(x) 
         S_u(x) + \gamma_5 S_u(x) \gamma_5 
         S_b^\prime(x) S_d(x) \nnb \\
\ar 
       S_d(x) S_b^\prime(x) \gamma_5 S_u(x) \gamma_5 + 
       S_u(x) S_b^\prime(x) \gamma_5 S_d(x) \gamma_5 + 
       \gamma_5 S_u(x) 
        Tr \left[S_b(x) \gamma_5 S_d^\prime(x) \right] \nnb \\
\ar 
       S_u(x) \gamma_5 
        Tr \left[S_b(x) S_d^\prime(x) \gamma_5 \right] + 
       \gamma_5 S_d(x) 
        Tr \left[S_u(x) \gamma_5 S_b^\prime(x) \right] + 
       S_d(x) \gamma_5 
        Tr \left[S_u(x) S_b^\prime(x) \gamma_5 \right] \Big) \nnb \\
\ar
      t^2  \Big( S_d(x) \gamma_5 S_b^\prime(x) \gamma_5 
         S_u(x) + S_u(x) \gamma_5 S_b^\prime(x) 
         \gamma_5 S_d(x) + 
       S_d(x) Tr \left[S_b(x) \gamma_5 S_u^\prime(x) 
          \gamma_5 \right] \nnb \\
\ar S_u(x) 
        Tr \left[S_d(x) \gamma_5 S_b^\prime(x) \gamma_5 \right] \Big) \Big\}~,\nnb 
\eea
where $S^\prime=CS^TC$, $T$ symbolizes transposition operator,
and $S$ is the quark (light or heavy) propagator. The corresponding
expressions of the correlation
functions for the other members of the sextet baryons can be found from
the $\Sigma_{B_Q}$ by making the following replacements:
\bea
\label{nolabel95}
\Pi^{\Sigma_c^0}         \es \Pi^{\Sigma_c^+}(u \rightarrow d)~, \nnb \\
\Pi^{\Sigma_c^{++}}      \es \Pi^{\Sigma_c^+}(d \rightarrow u)~, \nnb \\
\Pi^{\Sigma_b^-}         \es \Pi^{\Sigma_b^0}(u \rightarrow d)~, \nnb \\
\Pi^{\Sigma_b^+}         \es \Pi^{\Sigma_b^0}(d \rightarrow u)~, \nnb \\
\Pi^{\Xi_c^{\prime +}}   \es \Pi^{\Sigma_c^+}(d \rightarrow s)~, \nnb \\
\Pi^{\Xi_c^{\prime 0}}   \es \Pi^{\Sigma_c^+}(u \rightarrow d,d \rightarrow s)~, \nnb \\    
\Pi^{\Xi_b^{\prime 0}}   \es \Pi^{\Sigma_b^0}(d \rightarrow s)~, \nnb \\
\Pi^{\Xi_b^{\prime -}}   \es \Pi^{\Sigma_b^0}(u \rightarrow d,d \rightarrow s)~, \nnb \\    
\Pi^{\Omega_c^{0}} \es \Pi^{\Sigma_c^+}(u \rightarrow s,d \rightarrow s)~, \nnb \\
\Pi^{\Omega_b^{-}} \es \Pi^{\Sigma_b^0}(u \rightarrow s,d \rightarrow s)\nnb~.
\eea

The correlation function described by Eq. (\ref{enil09}) contains three
different parts: a) Perturbative part, when a photon is radiated from short
distance (perturbative contribution); b) a photon is radiated from short
distance from the quark propagators, and light
quarks form a condensate; c) a photon is radiated from long distance
(nonperturbative contribution).

In order to calculate the perturbative contribution of the correlation
function, it is enough to make the replacement
\bea
\label{nolabel01}
S \to \int d^4y S^{free}(x-y) \not\!\!{A} (y) S^{free}(y)~, \nnb
\eea
for one of the quark propagator,
where $S_{free}$ is the free quark operator; and the other two are being the
free propagators. The expressions of the free light
and heavy quarks in coordinate representation are given as,
\bea
\label{nolabel02}
S_q^{free} \es {i \!\!\not\!{x} \over 2 \pi^2 x^4} - {m_q \over 4 \pi^2 x^2}~, \nnb \\
S_Q^{free} \es {m_{B_Q}^2 \over 4 \pi^2} {K_1 (m_{B_Q}\sqrt{-x^2}) \over
\sqrt{-x^2} } + i {m_{B_Q}^2 \over 4 \pi^2} {\not\!{x} K_2 (m_{B_Q}\sqrt{-x^2}) \over
(\sqrt{-x^2})^2 }~, \nnb
\eea
where $K_i$ are the Bessel functions.

The contribution of part (c) can easily be calculated by replacing one of the
light quark operator with,
\bea
\label{nolabel03}
\left(S_q^{ab}\right)_{\alpha\beta} \to - {1\over 4} \delta^{ab} \bar{q}^a
\Gamma_i q^b \left(\Gamma_i\right)_{\alpha\beta}~,\nnb
\eea
where $\Gamma_j$ are the full set of Dirac matrices, and other quark
operators that involve perturbative, as well as nonperturbative terms.
The explicit forms of the ``full" quark propagators can be found in
\cite{Rnil19} and \cite{Rnil20}.

Nonperturbative contribution is realized as the matrix element of the
nonlocal operator $\bar{q} \Gamma_i q$ between the vacuum and one-photon
states.These matrix elements are described in terms of photon distribution
amplitudes, as are given below (see \cite{Rnil21}):

\bea
\label{nolabel04}
&&\langle \gamma(q) \vert  \bar q(x) \sigma_{\mu \nu} q(0) \vert  0
\rangle  = -i e_q \bar q q (\varepsilon_\mu q_\nu - \varepsilon_\nu
q_\mu) \int_0^1 du e^{i \bar u qx} \left(\chi \varphi_\gamma(u) +
\frac{x^2}{16} \mathbb{A}  (u) \right) \nnb \\ &&
-\frac{i}{2(qx)}  e_q \qq \left[x_\nu \left(\varepsilon_\mu - q_\mu
\frac{\varepsilon x}{qx}\right) - x_\mu \left(\varepsilon_\nu -
q_\nu \frac{\varepsilon x}{q x}\right) \right] \int_0^1 du e^{i \bar
u q x} h_\gamma(u)
\nnb \\
&&\langle \gamma(q) \vert  \bar q(x) \gamma_\mu q(0) \vert 0 \rangle
= e_q f_{3 \gamma} \left(\varepsilon_\mu - q_\mu \frac{\varepsilon
x}{q x} \right) \int_0^1 du e^{i \bar u q x} \psi^v(u)
\nnb \\
&&\langle \gamma(q) \vert \bar q(x) \gamma_\mu \gamma_5 q(0) \vert 0
\rangle  = - \frac{1}{4} e_q f_{3 \gamma} \epsilon_{\mu \nu \alpha
\beta } \varepsilon^\nu q^\alpha x^\beta \int_0^1 du e^{i \bar u q
x} \psi^a(u)
\nnb \\
&&\langle \gamma(q) | \bar q(x) g_s G_{\mu \nu} (v x) q(0) \vert 0
\rangle = -i e_q \qq \left(\varepsilon_\mu q_\nu - \varepsilon_\nu
q_\mu \right) \int {\cal D}\alpha_i e^{i (\alpha_{\bar q} + v
\alpha_g) q x} {\cal S}(\alpha_i)
\nnb \\
&&\langle \gamma(q) | \bar q(x) g_s \tilde G_{\mu \nu} i \gamma_5 (v
x) q(0) \vert 0 \rangle = -i e_q \qq \left(\varepsilon_\mu q_\nu -
\varepsilon_\nu q_\mu \right) \int {\cal D}\alpha_i e^{i
(\alpha_{\bar q} + v \alpha_g) q x} \tilde {\cal S}(\alpha_i)
\nnb \\
&&\langle \gamma(q) \vert \bar q(x) g_s \tilde G_{\mu \nu}(v x)
\gamma_\alpha \gamma_5 q(0) \vert 0 \rangle = e_q f_{3 \gamma}
q_\alpha (\varepsilon_\mu q_\nu - \varepsilon_\nu q_\mu) \int {\cal
D}\alpha_i e^{i (\alpha_{\bar q} + v \alpha_g) q x} {\cal
A}(\alpha_i)
\nnb \\
&&\langle \gamma(q) \vert \bar q(x) g_s G_{\mu \nu}(v x) i
\gamma_\alpha q(0) \vert 0 \rangle = e_q f_{3 \gamma} q_\alpha
(\varepsilon_\mu q_\nu - \varepsilon_\nu q_\mu) \int {\cal
D}\alpha_i e^{i (\alpha_{\bar q} + v \alpha_g) q x} {\cal
V}(\alpha_i) \nnb \\ && \langle \gamma(q) \vert \bar q(x)
\sigma_{\alpha \beta} g_s G_{\mu \nu}(v x) q(0) \vert 0 \rangle  =
e_q \qq \left\{
        \left[\left(\varepsilon_\mu - q_\mu \frac{\varepsilon x}{q x}\right)\left(g_{\alpha \nu} -
        \frac{1}{qx} (q_\alpha x_\nu + q_\nu x_\alpha)\right) \right. \right. q_\beta
\nnb \\ && -
         \left(\varepsilon_\mu - q_\mu \frac{\varepsilon x}{q x}\right)\left(g_{\beta \nu} -
        \frac{1}{qx} (q_\beta x_\nu + q_\nu x_\beta)\right) q_\alpha
\nnb \\ && -
         \left(\varepsilon_\nu - q_\nu \frac{\varepsilon x}{q x}\right)\left(g_{\alpha \mu} -
        \frac{1}{qx} (q_\alpha x_\mu + q_\mu x_\alpha)\right) q_\beta
\nnb \\ &&+
         \left. \left(\varepsilon_\nu - q_\nu \frac{\varepsilon x}{q.x}\right)\left( g_{\beta \mu} -
        \frac{1}{qx} (q_\beta x_\mu + q_\mu x_\beta)\right) q_\alpha \right]
   \int {\cal D}\alpha_i e^{i (\alpha_{\bar q} + v \alpha_g) qx} {\cal T}_1(\alpha_i)
\nnb \\ &&+
        \left[\left(\varepsilon_\alpha - q_\alpha \frac{\varepsilon x}{qx}\right)
        \left(g_{\mu \beta} - \frac{1}{qx}(q_\mu x_\beta + q_\beta x_\mu)\right) \right. q_\nu
\nnb \\ &&-
         \left(\varepsilon_\alpha - q_\alpha \frac{\varepsilon x}{qx}\right)
        \left(g_{\nu \beta} - \frac{1}{qx}(q_\nu x_\beta + q_\beta x_\nu)\right)  q_\mu
\nnb \\ && -
         \left(\varepsilon_\beta - q_\beta \frac{\varepsilon x}{qx}\right)
        \left(g_{\mu \alpha} - \frac{1}{qx}(q_\mu x_\alpha + q_\alpha x_\mu)\right) q_\nu
\nnb \\ &&+
         \left. \left(\varepsilon_\beta - q_\beta \frac{\varepsilon x}{qx}\right)
        \left(g_{\nu \alpha} - \frac{1}{qx}(q_\nu x_\alpha + q_\alpha x_\nu) \right) q_\mu
        \right]
    \int {\cal D} \alpha_i e^{i (\alpha_{\bar q} + v \alpha_g) qx} {\cal T}_2(\alpha_i)
\nnb \\ &&+
        \frac{1}{qx} (q_\mu x_\nu - q_\nu x_\mu)
        (\varepsilon_\alpha q_\beta - \varepsilon_\beta q_\alpha)
    \int {\cal D} \alpha_i e^{i (\alpha_{\bar q} + v \alpha_g) qx} {\cal T}_3(\alpha_i)
\nnb \\ &&+
        \left. \frac{1}{qx} (q_\alpha x_\beta - q_\beta x_\alpha)
        (\varepsilon_\mu q_\nu - \varepsilon_\nu q_\mu)
    \int {\cal D} \alpha_i e^{i (\alpha_{\bar q} + v \alpha_g) qx} {\cal T}_4(\alpha_i)
                        \right\}. \nnb
\eea
In the definitions given above, $\chi$ is the magnetic susceptibility of the
quarks,  
$\varphi_\gamma(u)$ is the leading twist-2, $\psi^v(u)$,
$\psi^a(u)$, ${\cal A}$ and ${\cal V}$ are the twist-3, and
$h_\gamma(u)$, $\mathbb{A}$, ${\cal T}_i$ ($i=1,~2,~3,~4$) are the
twist-4 photon DAs, respectively, whose explicit expressions are given in
Appendix A. 
The measure ${\cal D} \alpha_i$ is defined as
\bea
\label{nolabel05}
\int {\cal D} \alpha_i = \int_0^1 d \alpha_{\bar q} \int_0^1 d
\alpha_q \int_0^1 d \alpha_g \delta(1-\alpha_{\bar
q}-\alpha_q-\alpha_g).\nnb
\eea

In constructing the sum rules for the magnetic moments of heavy sextet
baryons it is necessary to equate the coefficients of the structure
$\rlap/{p}\rlap/{\varepsilon}\rlap/{q}$ to from the phenomenological and
theoretical representations of the correlation function. The following
steps in obtaining the final form of the sum rules for the magnetic moment
are: Fourier transformation, Borel transformation over the variables $p^2$
and $(p+q)^2$ variables, and continuum subtraction in order to suppress the
contribution of the higher states and continuum. After these procedures we
get the sum rules for the magnetic moment of the sextet heavy baryons, which
can schematically be written in the following form,
\bea
\label{nolabel06}
\lambda_{B_{Q_i}}^2 \mu_{B_{Q_i}} e^{m_{B_{Q_i}}^2/M^2} =
\Pi^{B,theor}~, \nnb
\eea
where the function $\Pi^{B,theor}$ contain perturbative and
nonperturbative contributions.

The expression for $\Pi^{B,theor}$ is quite lengthy, and for this reason we
do not present it here.

In order to determine the magnetic moments of heavy sextet baryons, the value of the
overlap amplitude is needed, which can be found from the analysis of the
two-point correlation function. The residues of the heavy sextet baryons are
calculated in \cite{Rnil22}, and their expressions can be found in this work.

\section{Results and discussion}

Having obtained the sum rules for the magnetic moments of the heavy sextet
spin-1/2 baryons, we are now ready to perform the numerical analysis.
The main input parameters of the light cone QCD sum rules for the magnetic
moments are the photon distribution functions DAs \cite{Rnil21}. Their
explicit expressions are presented in Appendix A. The values of the other
input parameters are: $\uu_{\mu=1 GeV} = \dd_{\mu=1 GeV} =
-(0.243)^3~GeV^3$, $\sp_{\mu=1 GeV} = 0.8 \uu_{\mu=1 GeV}$ \cite{Rnil23},
$m_0^2=(0.8\pm0.2)~GeV^2$ \cite{Rnil22}, $\Lambda = (0.5\pm0.1)~GeV$ \cite{Rnil24}
$f_{3\gamma}=-0.0039~GeV^2$ \cite{Rnil21},
${m_s}_{(\mu=2 GeV)}=(111\pm6)~MeV$ \cite{Rnil25}. 

Few words about the magnetic susceptibility involved in the numerical
analysis are in order. Its value is estimated to have the
value $\chi(1~GeV)=-4.4~GeV^{-2}$ in \cite{Rnil26}. In \cite{Rnil21},
using the vector dominance model ansatz and QCD sum rules its value is
predicted to be $\chi(1~GeV)=-(3.15 \pm 0.15)~GeV^{-2}$. Furthermore, from
an analysis of the radiative decays of heavy mesons its value is found to be
$\chi(1~GeV)=-(2.85 \pm 0.50)~GeV^{-2}$ \cite{Rnil27}. In numerical analysis
we have used all these predicted values of the magnetic susceptibility.

It should also be remembered that, the sum rules for the magnetic moments
involve the following auxiliary parameters: The arbitrary parameter $t$
appearing in the expressions of the interpolating currents, Borel mass
parameter $M^2$, and the continuum threshold $s_0$. The magnetic moment must
be independent of these parameters. In order to find ``regions" of these
parameters where magnetic moment exhibits good stability with respect to
their variations we proceed as follows. Firstly, we try to find the upper
and lower bounds of Borel mass parameter at fixed values of $s_0$ and $t$.
The upper bound of $M^2$ can be determined by requiring that the
contribution due to the continuum  threshold should be less than half of the
contribution coming from the perturbative part. The lower bound of $M^2$ can
be found by demanding that the higher powers of $1/M^2$ contribute less than
the leading twist contributions. Our numerical analysis shows that these two
conditions are satisfied simultaneously if $M^2$ ranges in the following
regions:
\bea
\label{nolabel09}
&& 2.0~GeV^2 \le M^2 \le 3.0~GeV^2~~~\Sigma_c\nnb\\ 
&& 2.2~GeV^2 \le M^2 \le 3.4~GeV^2~~~\Xi_c^\prime\nnb\\
&& 2.5~GeV^2 \le M^2 \le 4.0~GeV^2~~~\Omega_c\nnb\\
&& 5.0~GeV^2 \le M^2 \le 6.0~GeV^2~~~\Sigma_b\nnb\\
&& 5.0~GeV^2 \le M^2 \le 6.4~GeV^2~~~\Xi_b^\prime\nnb\\
&& 5.2~GeV^2 \le M^2 \le 7.0~GeV^2~~~\Omega_b\nnb
\eea

In regard to the continuum threshold $s_0$ appearing in the sum rules,
it is known that the values of this arbitrary
parameter is related to the energy of first excited state. The difference
$\sqrt{s_0}-m_{ground}$, where $m_{ground}$ is the ground state mass of the
baryon, is the energy needed to excite the particle to its first energy
state. This quantity usually varies between $0.3~GeV$ and $0.8~GeV$.
Analysis of the mass sum rules shows that, in order to reproduce the
experimental mass of the sextet baryons the continuum threshold should vary
in the following domain:
\bea
\label{nolabel10}
\sqrt{s_0} = \left\{ \begin{array}{c}
(3.1\pm 0.1)~GeV~~~\Sigma_c \\
(3.2\pm 0.1)~GeV~~~\Xi_c^\prime \\
(3.3\pm 0.1)~GeV~~~\Omega_c \\
(6.6\pm 0.2)~GeV~~~\Sigma_b \\     
(6.7\pm 0.2)~GeV~~~\Xi_b^\prime \\ 
(6.8\pm 0.2)~GeV~~~\Omega_b
\end{array} \right.
\eea

Having determined the sum rules, input parameters, and working regions of
all auxiliary parameters, we perform numerical analysis to calculate the
magnetic moment $\mu_{B_Q}$ of the  heavy sextet spin-1/2 baryons, whose results we can
summarize as follows.
As examples, in Figs. 1 and 2 we present the dependence of $\mu_{\Sigma_Q}$ on $M^2$ at
several fixed values of $t$ and at the fixed value of $s_0$ listed above for
the baryons $\Sigma_c^{++}$ and 
$\Sigma_b^{0}$, respectively. We observe
from these figures that the magnetic moments of $\Sigma_Q$ show good
stability with respect to the variation of the Borel mass parameter in
aforementioned domains. The next step in finding the values of the magnetic
moments of the baryons under consideration is to determine working region of
the arbitrary parameter $t$, where $\mu_{B_Q}$ is practically independent of
its variation. For this purpose, in Figs. 3 and 4 the dependence of
$\mu_{\Sigma_Q}$ on $\cos\theta$, where $\tan\theta=t$ for the
$\Sigma_c^{++}$ and $\Sigma_b^{0}$ baryons, at several fixed
values of the Borel mass parameter $M^2$ and at the predetermined value of
the continuum threshold $s_0$, are presented, respectively. It follows from these figures
that the magnetic moments of the heavy sextet spin-1/2 baryons seem to be
practically independent of the arbitrary parameter $t$ in the domain $-0.9
\le \cos\theta \le -0.6$, and insensitive to
the the chosen values of $s_0$. Our final results on the magnetic moments of
the heavy sextet spin-1/2 baryons are presented in Table 2. For comparison, in
the same Table we present the predictions of other approaches on the
magnetic moments of the relevant baryons, such as nonrelativistic quark
model \cite{Rnil11}, bag model \cite{Rnil09}, phenomenological relativistic
quark model \cite{Rnil12}, quark model with confinement law potential
\cite{Rnil13}, relativistic quark model \cite{Rnil04}, hypercentral model
\cite{Rnil06}, chiral constituent model \cite{Rnil14}, and QCD sum rules
method \cite{Rnil10}.    
    
\begin{table}[h]

\renewcommand{\arraystretch}{1.3}
\addtolength{\arraycolsep}{-0.5pt}
\small
$$
\begin{array}{|l|c|c|c|c|c|c|c|c|c|}
\hline \hline   
 &  \cite{Rnil04} & \cite{Rnil06} & \cite{Rnil09} & \cite{Rnil10} &
    \cite{Rnil11} &\cite{Rnil12}  & \cite{Rnil13} & \cite{Rnil14} & 
    \mbox{Our Work} \\
\hline \hline
\Sigma_c^{0}     & -1.04  & -1.015 & -1.043 & -1.6 \pm 0.2 & -1.37 &
                   -1.38 & -1.39\phantom{0}  & -1.6\phantom{0}  & -1.5    \\
\Sigma_c^{+}     &  \phantom{-}0.36  &  \phantom{-}0.5\phantom{00} & \phantom{-}0.318 &
\phantom{-}0.6 \pm 0.1 &  \phantom{-}0.49 &   
                    \phantom{-}0.49   &
\phantom{-}0.525   & \phantom{-}0.3\phantom{0}  &
\phantom{-}0.5 \\
\Sigma_c^{++}    &  \phantom{-}1.76  &  \phantom{-}2.279 &  \phantom{-}1.679 &
\phantom{-}2.1 \pm 0.3  &  \phantom{-}2.35 &   
                    \phantom{-}2.36 &  \phantom{-}2.44\phantom{0}  & \phantom{-}2.22  &
\phantom{-}2.4 \\
\Xi_c^{\prime 0} & -0.95  & -0.966 & -0.914 & \mbox{--} & -1.18 &   
                   -1.12 & -1.12\phantom{0}  &  -1.32    & -1.2 \\ 
\Xi_c^{\prime +} &  \phantom{-}0.47  &  \phantom{-}0.711 &  \phantom{-}0.591 &
\mbox{--}  &  \phantom{-}0.89 &  
                    \phantom{-}0.75 &  \phantom{-}0.796  & \phantom{-}0.76    &
\phantom{-}0.8 \\ 
\Omega_c^{0}     & -0.85  & -0.96\phantom{0} & -0.774 & \mbox{--} & -0.94 &  
                   -0.86  & -0.85\phantom{0}   & -0.9\phantom{0}    & -0.9 \\
\hline \hline
\Sigma_b^{-}     & -1.01  & -1.047 & -0.778 & \mbox{--} & -1.22 &
                   \mbox{--} & -1.256   & \mbox{--}  & -1.3    \\
\Sigma_b^{0}     &  \phantom{-}0.53  &  \phantom{-}0.592 & \phantom{-}0.422 &
\mbox{--} &  \phantom{-}0.64 &   
                    \mbox{--} & \phantom{-}0.659    & \mbox{--}  &
\phantom{-}0.6 \\
\Sigma_b^{+}     &  \phantom{-}2.07   &  \phantom{-}2.229 & \phantom{-}1.622 &
\mbox{--} &  \phantom{-}2.5\phantom{0} &   
                    \mbox{--} & \phantom{-}2.575  &  \mbox{--}  &
\phantom{-}2.4 \\
\Xi_b^{\prime -} & -0.91  & -0.902  & -0.66\phantom{0} & \mbox{--} & -1.02 &   
                   \mbox{--} & -0.985  &  \mbox{--}    & -1.2 \\ 
\Xi_b^{\prime 0} &  \phantom{-}0.66   &  \phantom{-}0.766 & \phantom{0}0.556 &
\mbox{--} &  \phantom{-}0.9\phantom{0} &  
                    \mbox{--} & \phantom{-}0.930  &  \mbox{--}    &
\phantom{-}0.7 \\ 
\Omega_b^{-}     & -0.82  & -0.96\phantom{0} & -0.545 & \mbox{--} & -0.79 &  
                   \mbox{--}  & -0.714   &  \mbox{--}    & -0.8 \\
\hline \hline
\end{array}
$$
\caption{Magnetic moments of the heavy sextet, $J^P = {1\over 2}^+$
baryons in units of the nuclear magneton $\mu_N$.}
\renewcommand{\arraystretch}{1}
\addtolength{\arraycolsep}{-1.0pt}
\end{table}

We observe from this table that almost all approaches give, more or less,
similar predictions, except the results of \cite{Rnil09} and
\cite{Rnil04} (especially for the charmed baryons), which are smaller.

Using our results one can easily deduce the following relations among the
magnetic moments of heavy baryons.
\bea
\label{nolabel11}
\mu_{\Sigma_c^{++}} + \mu_{\Sigma_c^{0}} \se 2 \mu_{\Sigma_c^{+}}~,\nnb \\
\mu_{\Sigma_c^{++}} + \mu_{\Omega_c^{0}} \se 2 \mu_{\Xi_c^{\prime +}}~,\nnb \\
\mu_{\Sigma_c^{++}} + 2 \mu_{\Xi_c^{\prime 0}} \se \mu_{\Sigma_c^{0}} + 2
\mu_{\Xi_c^{\prime +}}~,\nnb \\
\mu_{\Sigma_b^{+}} + \mu_{\Sigma_b^{-}} \se  2 \mu_{\Sigma_b^{0}}~,\nnb \\   
\mu_{\Sigma_b^{+}} + \mu_{\Omega_b^{-}} \se 2 \mu_{\Xi_b^{\prime 0}}~,\nnb \\
\mu_{\Sigma_b^{+}} + 2 \mu_{\Xi_b^{\prime -}} \se \mu_{\Sigma_b^{-}} + 2
\mu_{\Xi_b^{\prime 0}}~,\nnb
\eea

The direct measurement of the magnetic moments of heavy baryons are unlikely
in the near future. Therefore, any indirect estimations of the magnetic
moments of the heavy baryons could be very useful. For example, it could
help extracting information about the mass spectrum of the heavy baryons.
As is well known, experimentally measured mass difference is attributed to
the hyperfine splitting. Moreover, the magnetic moments of quarks are
proportional to the chromomagnetic moments, which determine the hyperfine
splitting in the baryon spectrum. Following this reasoning, the magnetic
moments of the $\Lambda_c$ and $\Lambda_b$ baryons are estimated in
\cite{Rnil28}. Hence, this approach could be an essential tool in
estimating the magnetic moments of the heavy baryons.

In conclusion, the magnetic moments of the heavy sextet $J^P = {1\over
2}^+$ baryons are calculated in framework of the light cone QCD sum rules
method. Empirically, linearly independent relations among the magnetic
moments of the sextet baryons are obtained. Comparison of our results with
the predictions of other approaches is presented.

\newpage


\section*{Appendix A: Photon distribution amplitudes}
\setcounter{equation}{0}
\setcounter{section}{0}


Explicit forms of the photon DAs
\cite{Rnil21}.

\bea
\label{nolabel27}
\varphi_\gamma(u) \es 6 u \bar u \Big[ 1 + \varphi_2(\mu)
C_2^{\frac{3}{2}}(u - \bar u) \Big]~,
\nnb \\
\psi^v(u) \es 3 [3 (2 u - 1)^2 -1 ]+\frac{3}{64} (15
w^V_\gamma - 5 w^A_\gamma)
                        [3 - 30 (2 u - 1)^2 + 35 (2 u -1)^4]~,
\nnb \\
\psi^a(u) \es [1- (2 u -1)^2] [ 5 (2 u -1)^2 -1 ]
\frac{5}{2}
    \Bigg(1 + \frac{9}{16} w^V_\gamma - \frac{3}{16} w^A_\gamma
    \Bigg)~,
\nnb \\
{\cal A}(\alpha_i) \es 360 \alpha_q \alpha_{\bar q} \alpha_g^2
        \Bigg[ 1 + w^A_\gamma \frac{1}{2} (7 \alpha_g - 3)\Bigg]~,
\nnb \\
{\cal V}(\alpha_i) \es 540 w^V_\gamma (\alpha_q - \alpha_{\bar q})
\alpha_q \alpha_{\bar q}
                \alpha_g^2~,
\nnb \\
h_\gamma(u) \es - 10 (1 + 2 \kappa^+ ) C_2^{\frac{1}{2}}(u
- \bar u)~,
\nnb \\
\mathbb{A}(u) \es 40 u^2 \bar u^2 (3 \kappa - \kappa^+ +1 ) +
        8 (\zeta_2^+ - 3 \zeta_2) [u \bar u (2 + 13 u \bar u) + 
                2 u^3 (10 -15 u + 6 u^2) \ln(u) \nnb \\ 
\ar 2 \bar u^3 (10 - 15 \bar u + 6 \bar u^2)
        \ln(\bar u) ]~,
\nnb \\
{\cal T}_1(\alpha_i) \es -120 (3 \zeta_2 + \zeta_2^+)(\alpha_{\bar
q} - \alpha_q)
        \alpha_{\bar q} \alpha_q \alpha_g~,
\nnb \\
{\cal T}_2(\alpha_i) \es 30 \alpha_g^2 (\alpha_{\bar q} - \alpha_q)
    [(\kappa - \kappa^+) + (\zeta_1 - \zeta_1^+)(1 - 2\alpha_g) +
    \zeta_2 (3 - 4 \alpha_g)]~,
\nnb \\
{\cal T}_3(\alpha_i) \es - 120 (3 \zeta_2 - \zeta_2^+)(\alpha_{\bar
q} -\alpha_q)
        \alpha_{\bar q} \alpha_q \alpha_g~,
\nnb \\
{\cal T}_4(\alpha_i) \es 30 \alpha_g^2 (\alpha_{\bar q} - \alpha_q)
    [(\kappa + \kappa^+) + (\zeta_1 + \zeta_1^+)(1 - 2\alpha_g) +
    \zeta_2 (3 - 4 \alpha_g)]~,\nnb \\
{\cal S}(\alpha_i) \es 30\alpha_g^2\{(\kappa +
\kappa^+)(1-\alpha_g)+(\zeta_1 + \zeta_1^+)(1 - \alpha_g)(1 -
2\alpha_g)\nnb \\ 
\ar\zeta_2
[3 (\alpha_{\bar q} - \alpha_q)^2-\alpha_g(1 - \alpha_g)]\}~,\nnb \\
\widetilde {\cal S}(\alpha_i) \es-30\alpha_g^2\{(\kappa -
\kappa^+)(1-\alpha_g)+(\zeta_1 - \zeta_1^+)(1 - \alpha_g)(1 -
2\alpha_g)\nnb \\ 
\ar\zeta_2 [3 (\alpha_{\bar q} -
\alpha_q)^2-\alpha_g(1 - \alpha_g)]\}. \nnb
\eea
The parameters entering  the above DA's are borrowed from
\cite{Rnil21} whose values are $\varphi_2(1~GeV) = 0$, 
$w^V_\gamma = 3.8 \pm 1.8$, $w^A_\gamma = -2.1 \pm 1.0$, 
$\kappa = 0.2$, $\kappa^+ = 0$, $\zeta_1 = 0.4$, $\zeta_2 = 0.3$, 
$\zeta_1^+ = 0$, and $\zeta_2^+ = 0$.


\newpage

\section*{Figure captions}
{\bf Fig. (1)} The dependence of the magnetic moment of the $\Sigma_c^{++}$ 
baryon on $M^2$, at several fixed values of $t$, and at
$s_0=9.0~GeV^2$.\\\\
{\bf Fig. (2)} The same as Fig. (1), but for the $\Sigma_b^{0}$ baryon, at
$s_0=41.0~GeV^2$.\\\\
{\bf Fig. (3)} The dependence of the magnetic moment of the $\Sigma_c^{++}$
baryon on $\cos\theta$, at several fixed values of $M^2$, and at
$s_0=9.0~GeV^2$.\\\\
{\bf Fig. (4)} The same as Fig. (3), but for the $\Sigma_b^{0}$ baryon, at    
$s_0=41.0~GeV^2$.\\\\

\newpage

\begin{figure}
\vskip 3. cm
    \includegraphics{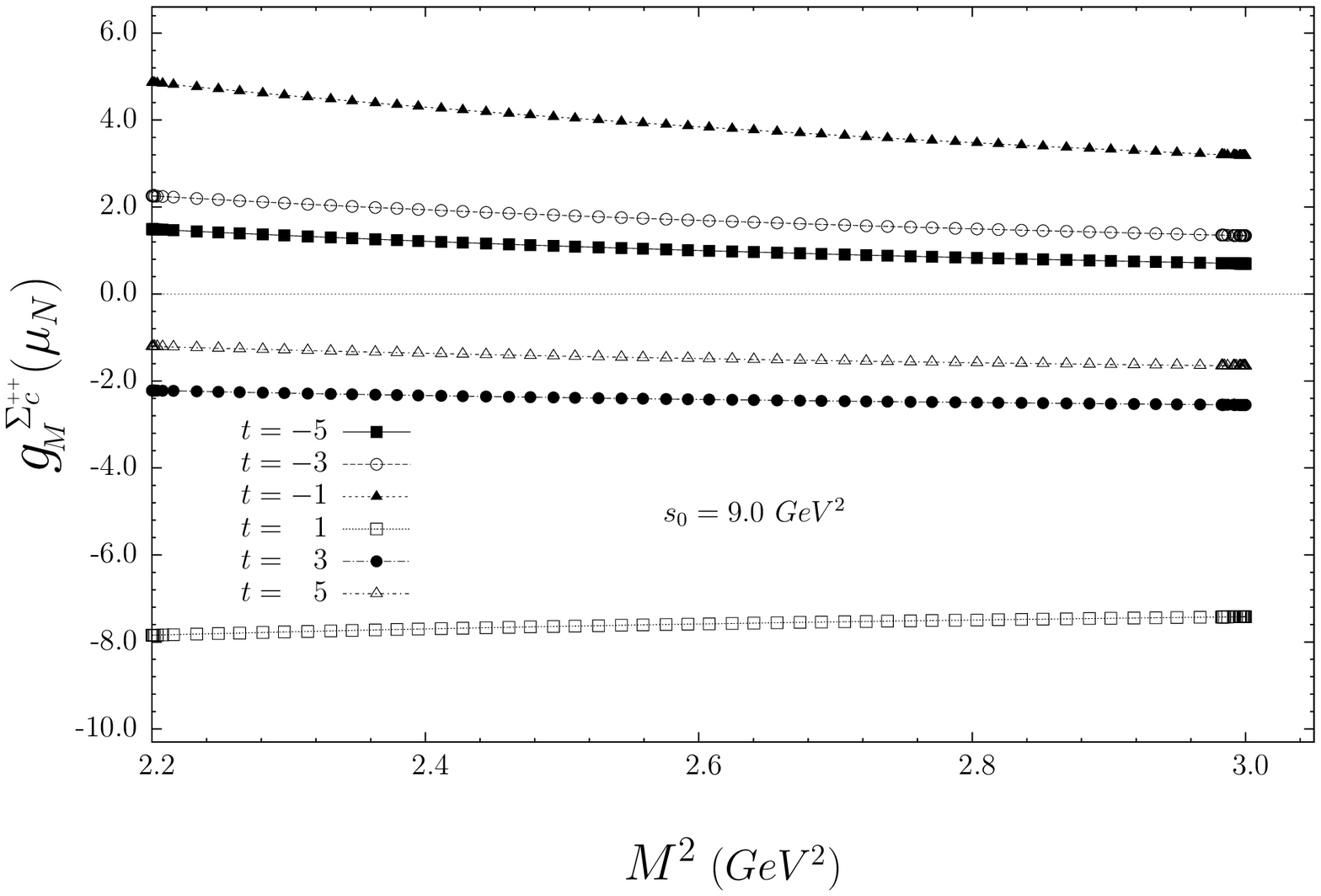}
\vskip 7.0cm
\caption{}
\end{figure}

\begin{figure}
\vskip 3. cm
    \includegraphics{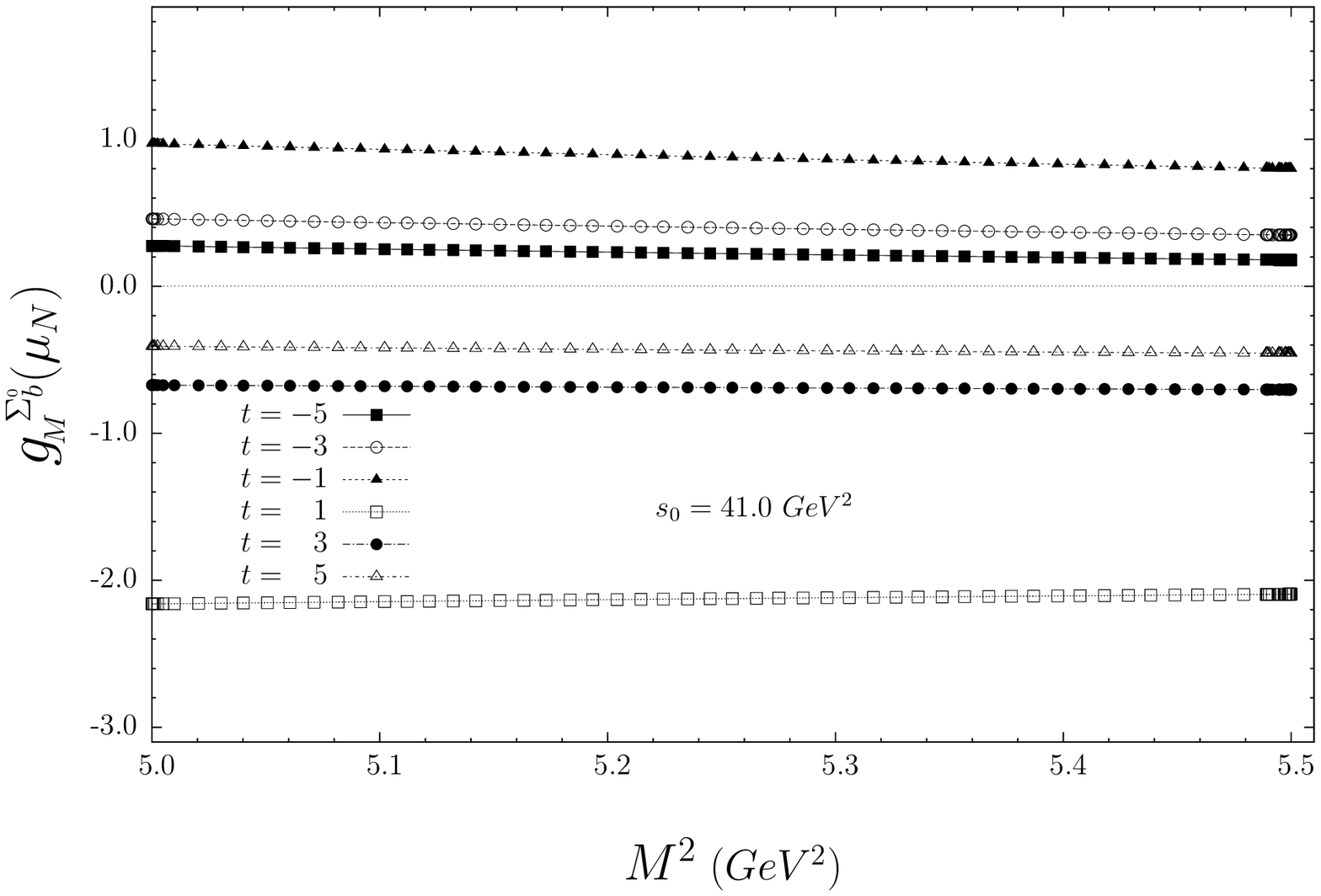}
\vskip 7.0cm
\caption{}
\end{figure}

\begin{figure}
\vskip 3. cm
    \includegraphics{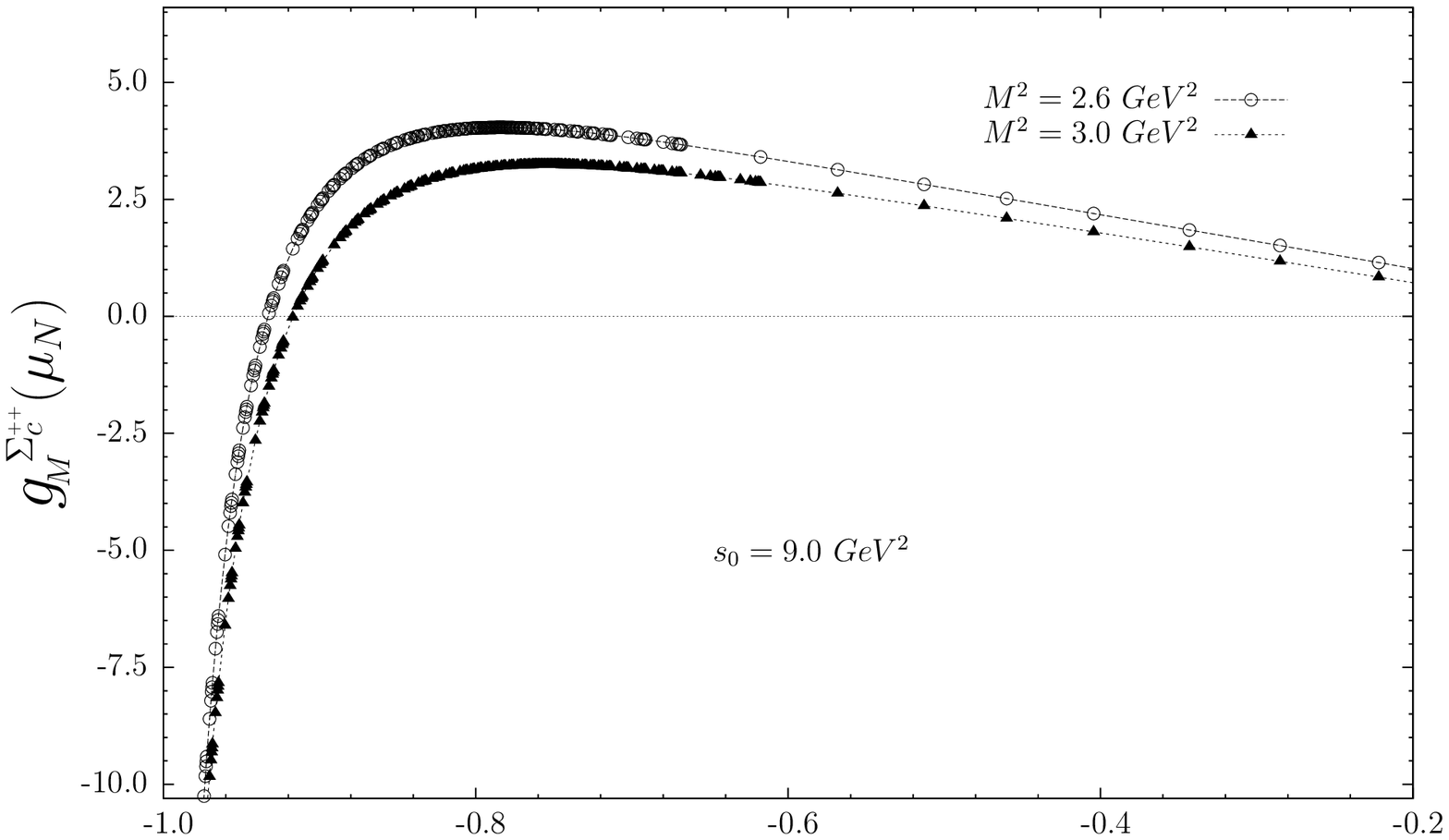}
\vskip 7.0cm
\caption{}
\end{figure}

\begin{figure}
\vskip 3. cm
    \includegraphics{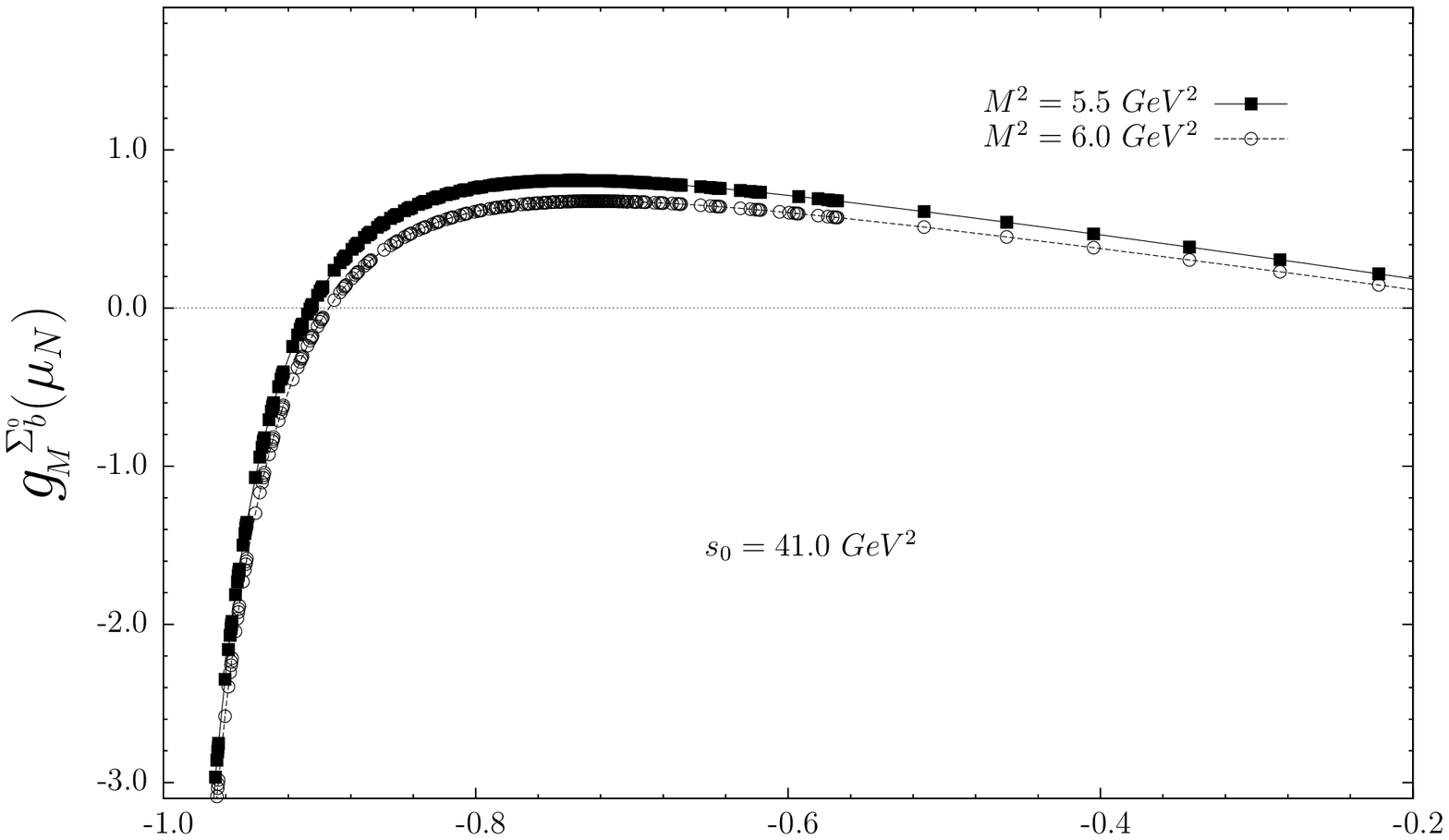}
\vskip 7.0cm
\caption{}
\end{figure}

\end{document}